\documentclass[letterpaper]{article} % DO NOT CHANGE THIS
\usepackage{aaai25}  % DO NOT CHANGE THIS
\usepackage{times}  % DO NOT CHANGE THIS
\usepackage{helvet}  % DO NOT CHANGE THIS
\usepackage{courier}  % DO NOT CHANGE THIS
\usepackage[hyphens]{url}  % DO NOT CHANGE THIS
\usepackage{graphicx} % DO NOT CHANGE THIS
\usepackage{natbib}  % DO NOT CHANGE THIS AND DO NOT ADD ANY OPTIONS TO IT
\usepackage{caption} % DO NOT CHANGE THIS AND DO NOT ADD ANY OPTIONS TO IT
\usepackage{algorithm}
\usepackage{algorithmic}

\usepackage{newfloat}
\usepackage{listings}
\DeclareCaptionStyle{ruled}{labelfont=normalfont,labelsep=colon,strut=off} % DO NOT CHANGE THIS
\floatstyle{ruled}
\newfloat{listing}{tb}{lst}{}
\floatname{listing}{Listing}
\usepackage{booktabs}
\usepackage{graphicx}

\usepackage{xcolor}
\newcommand{\answerYes}[1]{\textcolor{blue}{#1}} 
\newcommand{\answerNo}[1]{\textcolor{teal}{#1}} 
\newcommand{\answerNA}[1]{\textcolor{gray}{#1}}

\title{Asking For It: Question-Answering for Predicting Rule Infractions in Online Content Moderation}
\author {
    Mattia Samory\textsuperscript{\rm 1},
    Diana Pamfile\textsuperscript{\rm 1},
    Andrew To\textsuperscript{\rm 2},
    Shruti Phadke\textsuperscript{\rm 2}
}
\affiliations {
    \textsuperscript{\rm 1}Sapienza University of Rome\\
    \textsuperscript{\rm 2}Drexel University\\
        mattia.samory@uniroma1.it, pamfile.1943337@studenti.uniroma1.it, dt686@drexel.edu, sp3945@drexel.edu
}
\usepackage{bibentry}
\begin{document}

\maketitle

\begin{abstract}
Online communities rely on a mix of platform policies and community-authored rules to define acceptable behavior and maintain order. However, these rules vary widely across communities, evolve over time, and are enforced inconsistently—posing challenges for transparency, governance, and automation. In this paper, we model the relationship between rules and their enforcement at scale, introducing ModQ, a novel question-answering framework for rule-sensitive content moderation. Unlike prior classification or generation-based approaches, ModQ conditions on the full set of community rules at inference time and identifies which rule best applies to a given comment. We implement two model variants—extractive and multiple-choice QA—and train them on large-scale datasets from Reddit and Lemmy, the latter of which we construct from publicly available moderation logs and rule descriptions. Both models outperform state-of-the-art baselines in identifying moderation-relevant rule violations, while remaining lightweight and interpretable. Notably, ModQ models generalize effectively to unseen communities and rules, supporting low-resource moderation settings and dynamic governance environments. 
\end{abstract}


\section{Introduction}

Content moderation on platforms like Reddit and Lemmy is often delegated to the community members, enabling decentralized governance structures where local norms, values, and enforcement thresholds reflect the collective will of each community in addition to platform-wide policies. This approach to moderation empowers communities with localized, context-specific decision-making where community members enforce their own standards of acceptable behavior, often through a set of community-specific rules. 

Local rules defined by communities are not static; they evolve dynamically in response to shifts in user demographics, emergent behaviors, platform-wide developments, and external sociopolitical contexts. Prior research has shown that such rules reflect the distinctive social, cultural, and organizational logics of the communities in which they are embedded, leading to significant variation even within a single platform ecosystem \cite{fiesler2018reddit, reddy2023evolution-daa, frey2022governing-dcb}. Moderation practices are thus highly contextual, temporally sensitive, and shaped by localized governance priorities.

Despite growing recognition of the sociotechnical complexity of moderation, most existing automated approaches treat rule enforcement as a binary classification task: should a given piece of content be removed or retained? These models primarily focus on the content itself—whether it constitutes, for instance, harassment or hate speech—rather than its alignment with community-specific rules. While recent studies have demonstrated that including rule-level information improves the predictive accuracy of moderation systems \cite{park2021detecting,xin2024let-503,he2024cpl}, these approaches are typically limited in scope. They assess content compliance one rule or category at a time and often require extensive computational resources, constraining their scalability and utility in real-world, volunteer-driven community settings.

To address these limitations, we introduce a novel formulation of community rule enforcement as an information extraction task. Leveraging a comprehensive, longitudinal dataset of moderation actions from Lemmy—a federated platform where rules and enforcement decisions are publicly logged— and existing Reddit moderation datasets, we model the relationship between user comments and the complete set of community rules. Rather than mapping content to a fixed label or rule category, our approach identifies the specific rule violated, conditioned on the full textual set of rules defined by the community at that point in time.

%However, contemporary computational approaches to modeling community moderation often miss out on learning from the rich, descriptive rules and regulations defined by individual communities. They often rely on generalized taxonomies, limited context or multiple data and time intensive classifiers 

%guided by localized, context specific rules that reflect the norms, values and governance styles of each community. Community rules are not static - they evolve in response to the growth of the community and 

%rules are contextual and change with time. Research showed how rules reflect key characteristics of the communities they are applied to, and that communities differ in how they moderate content. 
%\paragraph{Gap} However, little research accounts for these differences when developing and automating moderation tools.

%With few exceptions, most work has focused on the moderation action---whether to remove or keep a piece of content---or characteristics of the content itself---if it constitutes harassment---rather than community-specific rules. Recent works demonstrated how including rule information improves classifiers' ability to identify content to be removed according to each community's norms. However, these works were limited to specific rules and moderation cases, due to limitations in data access. Specifically, these models achieved state-of-the-art performance, but assessed a comment's compliance with one rule or rule category at a time. This approach requires significant computational resources, which limits the ability of communities to benefit from this technology. 

\begin{figure}[h]
    \centering
    \includegraphics[width=\columnwidth]{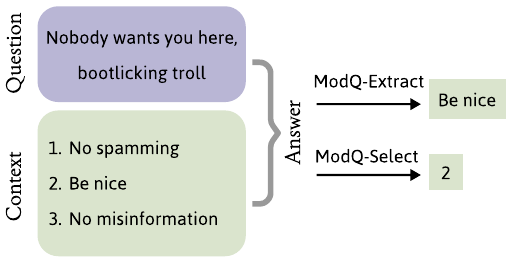}
    \caption{ModQ-Extract and ModQ-Select models presented in this paper for modeling content moderation as a question and answer task}
    \label{fig:introthumb}
\end{figure}

Specifically, we propose two novel information extraction models that consider moderation as a question-answering problem (Figure \ref{fig:introthumb})---ModQ-Extract, an extractive question answering (Q\&A) model, and ModQ-Select, a multiple-choice Q\&A model.
ModQ-Extract treats rule enforcement as a span prediction problem. Given a user comment and the full set of community rules as context, the ModQ-Extract is trained to extract the specific span corresponding to the rule that justifies moderation. ModQ-Select, by contrast, frames rule identification as a multiple-choice Q\&A task. For each comment, the model scores its alignment with each of the community’s rules and selects the most appropriate rule.  

We evaluate our proposed models against several strong baselines, including NormVio—a community-sensitive classifier for rule categories \cite{park2021detecting}—and CPL-NoViD \cite{he2024cpl}, a state-of-the-art prompting-based model that generates per-rule moderation decisions. Both of our proposed models, ModQ-Extract and ModQ-Select, demonstrate strong performance in identifying rule violations across Reddit (Table \ref{tab:reddit_test}) and Lemmy (Table \ref{tab:lemmy_test_stratified}) datasets. ModQ-Select consistently outperforms all baselines across rule categories and moderation tasks, while ModQ-Extract performs comparably to state-of-the-art methods despite addressing the more granular task of identifying the exact rule rather than a coarse rule category. Crucially, our models generalize better than existing approaches to unseen communities (Figure \ref{fig:n_communities_out}) and previously unencountered rules (Figure \ref{fig:n_rules_out}). This is a key advantage in federated or fast-evolving platforms like Lemmy, where new communities and rules are continually introduced. While performance predictably declines in these out-of-domain settings, both ModQ variants maintain a clear edge over prior models.

Overall, we offer a new formulation of rule-sensitive content moderation as a question-answering task, advancing both methodological and practical goals in the field of computational moderation. The key contributions are as follows:

\begin{itemize}
    \item  We introduce ModQ, a modeling approach that recasts moderation as either extractive or multiple-choice question answering. Unlike prior classification or generation-based methods, our framework is designed to be interpretable, computationally efficient, and capable of handling an open set of community-specific rules.
    \item By outputting rule-level predictions, the proposed models can be embedded in flagging systems, rule-hinting tools, or rationale-assistance features for moderators. This transparency enhances trust, reduces ambiguity in enforcement, and supports both moderators and end users.
    \item Our QA formulation can be used to simulate rule revisions, test alignment between stated and enacted policies, and better understand the institutional logics of community governance.
\end{itemize}

This work contributes to the ICWSM community bridging NLP and social computing to address real-world challenges in community-specific moderation, transparency, and platform governance at scale. 

%\paragraph{Novelty} We tackle these challenges by curating a novel dataset based on a complete longitudinal log of moderation actions on Lemmy. We propose a novel formulation that identifies which rule a comment infringes, given the entire set of community rules, before suggesting the appropriate moderation action. 
%\paragraph{Implications} This formulation allows, by design, the model to be applied even when rules change, and to learn across an open set of communities and rules. The model can be used for inference on commodity hardware, enabling community moderators to harness it.

% \paragraph{Contributions}   
% \begin{itemize}
%     \item We provide a novel formulation of automated moderation for community-specific rules.
%     \item We develop a reference implementation and show how it improves upon the state-of-the-art.
% \end{itemize}

\section{Related Work}

\subsection{Rules and Moderation Enforcement in Online Communities}
Given the increasing recognition of its impact in users' lives---from democratic processes to economic opportunities---the governance of online communities is increasingly subject to scrutiny. 
While legal and platform-wide regulatory frameworks establish parameters for user behavior \cite{katzenbach2023platform, vlist2022api-474}, the fine-grained norms and expectations are often articulated through community-specific rules, which vary significantly even within the same online platform \cite{fiesler2018reddit}. These rules are not static; rather, they evolve in response to the changing demographics, prevalent behaviors, and emergent issues within a community, reflecting its unique characteristics and temporal development \cite{reddy2023evolution-daa, frey2022governing-dcb}.

This time-varying and community-specific nature of rules poses a significant challenge for automated content moderation. Research has highlighted the importance of community context, including shared and idiosyncratic norms \cite{Chandrasekharan2018}; conversation context, where subtle cues can indicate violations \cite{Hardaker2013, Samory2017a}; user context, leveraging past behavior; interpretation context, addressing mismatches between moderators and users \cite{koshy2023measuring-4aa,munzert2025citizen-4a8}; and the difficulties of deliberation context in nuanced cases \cite{koshy2024venire-5f7}. The inability of current automated systems to fully capture and adapt to an increasing variety of community-specific rules, and to account for their evolving nature, remains a significant limitation.
%(e.g., TODOCITE)
This work directly addresses the gap by modeling the relationship between content and contextually-defined, community-specific rules. 

\subsection{Automated Content Moderation}
The increasing need for scalable solutions to online content moderation has bolstered the adoption of automated moderation systems that commonly employ rule-based filters \cite{jhaver2019human-machine-005}, keyword detection \cite{ribeiro2024post-2ce}, and sophisticated machine learning techniques to identify and flag potentially violating content \cite{chandrasekharan2019crossmod-516}. Natural Language Processing research contributed substantially to the advancement of the technology for detecting harmful language, such as hate speech and sexism \cite{yu2024unseen, Samory2021a}. The maturity of this field is evident in the commodification of automated moderation as a service (e.g., Jigsaw's Perspective APIs, OpenAI's Moderation API) \cite{parker2023is-86f}. However, a notable disconnection between NLP research and moderation practice is that the former often relies on definitions of undesirable content---and their operalizations in terms of datasets and tasks---which are often misaligned with the norms and requirements of online communities \cite{cao2024toxicity}. Thus, a growing body of research in the intersection of NLP and Social Computing focuses on developing resources and models that may directly support moderators \cite{koshy2024venire-5f7,chandrasekharan2019crossmod-516}. Closest to this paper, recent work aims to model the enforcement of community-specific rules \cite{park2021detecting,xin2024let-503,he2024cpl}.

The prevailing formulation of rule enforcement as an NLP task has been as a classification problem. In this paradigm, community rules are mapped to a predefined set of classes, and classifiers are trained to predict these classes for new content (e.g., \cite{park2021detecting,xin2024let-503}). One limitation of this approach is that rules are often mapped to coarse-grained taxonomies, despite the existence of thousands of distinct rules within platforms like Reddit. A second, more recently explored formulation involves language generation. Here, given a comment and contextual information, a model is tasked with generating the specific rule that applies (e.g., \cite{he2024cpl,wang2025end-4bf,zhan2024slm-mod-1bd}). This approach, leveraging sequence-to-sequence models or prompting large language models, offers the potential to model context-dependent moderation, as the same comment appearing in communities with different rules could theoretically be processed differently. However, it also suffers from a lack of inherent explainability and significant computational demands. The present work introduces a novel formulation of rule enforcement as an information extraction task, sharing the advantage of language generation in that we can model different rules for different communities, while relying on much more computationally efficient models. 

\subsection{Automated Governance beyond Sanctioning}

While automation offers scalability in managing vast quantities of user-generated content\cite{jhaver2019human-machine-005, wright2022automated-0d2}, it poses significant challenges \cite{gorwa2020algorithmic}. Especially, the lack of transparency in automated moderation raises concerns regarding over-censorship or the suppression of legitimate speech (e.g., \cite{thach2024visible}). Recent thrusts in social computing research identify compounded opportunities in developing interpretable automated moderation systems. Beyond the automation of sanctioning (e.g., content removal, user suspension), transparent automated moderation may support governance more broadly \cite{eslami2024future-c53, park2021detecting}. This includes leveraging computational tools to provide communities with insights into their norms, facilitate more informed rule-making processes, and support deliberative processes among community members and moderators (e.g., \cite{de2020modular,bajpai2024modeling-acb,kuo2024policycraft}).
The present work adds to this body of work toward developing transparent and interpretable moderation systems, to ultimately empower communities to take a more active and data-driven role in shaping their own online environments, moving beyond a purely reactive model of content policing (e.g., \cite{filippi2021editorial-24c}).

\begin{table*}[t]
\centering
\resizebox{\textwidth}{!}{%
\begin{tabular}{@{}rcccccl@{}}
\toprule
\multicolumn{1}{c}{} & \textbf{Modlogs} & \textbf{Communities} & \textbf{\begin{tabular}[c]{@{}c@{}}Avg. no. rules \\ per comment\end{tabular}} & \textbf{\begin{tabular}[c]{@{}c@{}}Safe \\ comments\end{tabular}} & \textbf{\begin{tabular}[c]{@{}c@{}}Removed \\ comments\end{tabular}} & \multicolumn{1}{c}{\textbf{\begin{tabular}[c]{@{}c@{}}Top 5 \\ communities\end{tabular}}} \\ \midrule
\textbf{Lemmy} & 40,534 & 413 & 6 & 21,518 & 19,016 & \begin{tabular}[c]{@{}l@{}}worldnews, memes, technology, \\ news, politics\end{tabular} \\ \midrule
\textbf{Reddit} & 50,732 & 2,264 & 3 & 31,119 & 19,613 & \begin{tabular}[c]{@{}l@{}}Coronavirus, AmItheAsshole, classicwow, \\ CanadaPolitics, Games\end{tabular} \\ \bottomrule
\end{tabular}%
}
\caption{Lemmy and Reddit data description}
\label{tab:main_data}
\end{table*}

\section{Data Collection and Curation}

We evaluate our approach using moderation data from two multi-community, peer-moderated platforms: Reddit and Lemmy. Tables \ref{tab:main_data} and  \ref{tab:cat_data} present summary statistics for both datasets.

\subsection{Reddit}
We leverage the Reddit moderation dataset introduced by \cite{park2021detecting}, which consists of 20K conversation threads in which the final comment was removed by a moderator. These moderated conversations were identified by locating moderator comments that explicitly cited a rule number or quoted rule text in a public response to the removed content. \citeauthor{park2021detecting} retrieved the entire preceding conversation thread—including the original post, all parent comments, and the removed comment itself—using the Pushshift API. Additionally, they included a control set of 32,000 unmoderated conversations, selected based on temporal proximity and matched to the same original post as their moderated counterparts. Each unmoderated conversation was associated with the rule applied in the corresponding moderated case. Finally, the rules were categorized according to the taxonomy developed by \citeauthor{fiesler2018reddit}, using BERT-based classifiers fine-tuned for each rule category.

While providing a valuable resource for modeling moderation, the dataset exhibits inherent limitations stemming from its collection methodology. By focusing exclusively on cases where moderators publicly explained their reasoning by referencing a rule, the dataset is limited to a specific subset of moderation actions --- those deemed by moderators to warrant such public justification, which may introduce a bias towards certain types of rule violations or moderation styles. In particular, the dataset only includes cases of the application of a select few rules in each community. We address this limitation by constructing a new dataset based on a full record of moderation actions and community rules.

To support our QA-based models, we augmented this dataset by identifying all instances of rules associated with each community across the original dataset. We compiled these rule mentions to construct a rule set for every community, which serves as the contextual input to our models. While the original dataset only includes a limited subset of applied rules, this reconstruction enables us to model rule selection in a more realistic setting where a more comprehensive set of community rules is available at inference time.
% \subsubsection{Building Reddit Data for Moderation Q\&A}
% [TODO - DID WE END UP USING ARCTIC SHIFT? No. we don't have to even have this section I think, if not for the exclusion of entries with multiple rules.]

\subsection{Lemmy} Lemmy, a popular platform in the Fediverse, presents itself as an open-source, non-commercial alternative to Reddit. Lemmy consists of thousands of federated instances that host volunteer-moderated communities.
We used Lemmy's public API to build a dataset of safe and removed comments. To discover communities on Lemmy, we gathered available lists of instances from multiple portals to the network (fedidb, fediverse observer, awesomelemmy, lemmyverse.net) and recursively snowballed through federated instances.

\begin{figure*}[]
    \centering
    \includegraphics[width=0.90\textwidth]{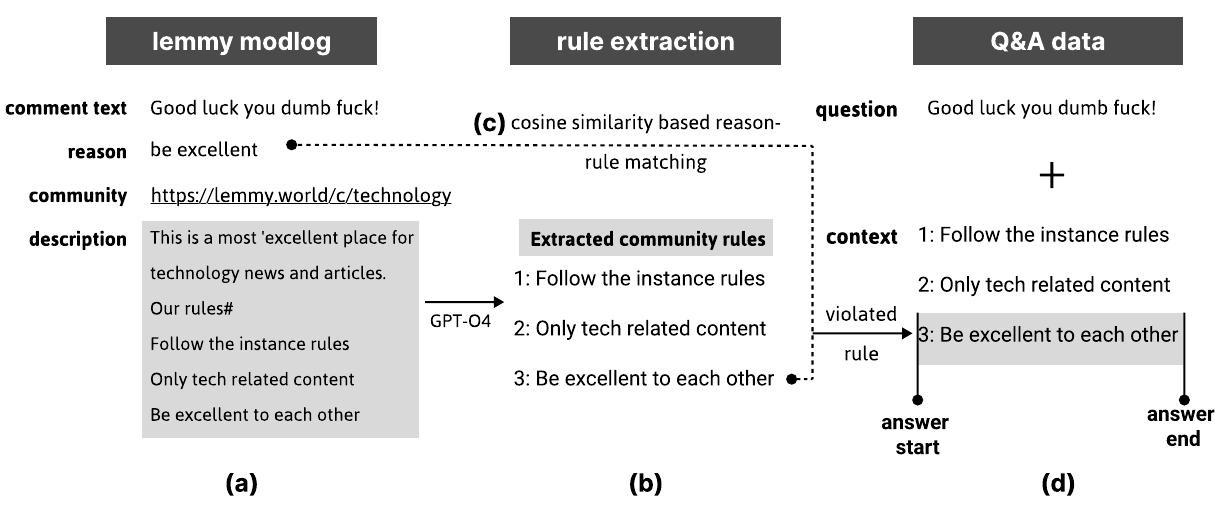}
    \caption{\textbf{Lemmy data preparation process: }Figure illustrating various stages in the data preparation phase for Lemmy modlogs. (a) displays a typical Lemmy modlog queried from Lemmy's public API. We use GPT-O4 mini to extract structured community rules from the community description (b) and match the removal \emph{reason} provided by moderators with one of the extracted rules (c). We then prepare Q\&A data for the bert model in the form of \emph{question}, \emph{context} and \emph{start and end} of the answer as displayed in (d). }
    \label{fig:lemmydata}
\end{figure*}

\paragraph{Moderated comments} Unlike most proprietary platforms, Lemmy provides direct access to moderation logs (modlogs) through its public and free API.\footnote{An example public modlog endpoint for Lemmy is \url{https://lemmy.world/api/v3/modlog}} Modlogs are a complete record of moderator actions, including content removals and rule enforcement justifications \cite{samory2021positive,li2022measuring-0b5,juneja2020through-eab} as displayed in Figure \ref{fig:lemmydata}(a). 

We collected moderator-removed comments from modlogs in communities local to each instance to avoid duplication. To maintain label accuracy, we excluded comments that had been mass-removed or later reinstated by moderators. We further restricted our analysis to comments that retained their original text, were in English, and included rule enforcement justifications provided by moderators.

\paragraph{Community rules} The free-text self-descriptions of communities on Lemmy serve as the conventional location for publishing community rules. We accessed community descriptions directly from modlogs at the time of each moderation action. This approach allowed us to retrieve the precise rule set in effect when the comment was removed, accounting for temporal changes in community governance.
We used GPT-4o to automate the structured extraction of rules from these descriptions (Figure~\ref{fig:lemmydata}(b)). To validate the accuracy of this extraction process, three authors manually evaluated 100 randomly sampled rule sets, finding a high accuracy rate of 86.9%. We limited our analysis to communities with no more than 20 rules, as communities exceeding this threshold were often used for testing and contained synthetic or non-substantive rule sets.

Following the approach used for Reddit, we categorized rules using classifiers fine-tuned on data from \cite{fiesler2018reddit}, mapping them to the coarser-grained categories used by \cite{park2021detecting, he2024cpl}. Rules that could not be mapped were grouped under the ``Other'' category.
It is important to note that the ModQ models proposed in this paper rely solely on the comment text and the extracted rule texts (Figure~\ref{fig:lemmydata}(d)); rule categories are not used during modeling. Rather, they are included solely to enable clearer comparisons with the baseline models.

\paragraph{Matching rules with removal reason} To identify the specific rule violated by a comment in the modlog, we begin by analyzing the \emph{reason} provided by moderators. In many cases, moderators explicitly cite a rule number (e.g., “Rule 6”) that corresponds directly to a rule in the community description—these references are captured using regular expressions.
However, moderators may also provide a free-text, descriptive rationale. To align such rationales with the correct rule, we tokenize both the moderator-provided reason and all extracted community rules using the lightweight \texttt{all-MiniLM} model.\footnote{\url{https://huggingface.co/sentence-transformers/all-MiniLM-L6-v2}} We then compute cosine similarity between the tokenized embeddings (Figure~\ref{fig:lemmydata}(c)). If the similarity between a reason and a rule exceeds a threshold of 0.85, we treat that rule as the match. While this threshold is heuristically chosen, manual inspection of 50 reason–rule mappings verified via similarity revealed no mismatches. In the final dataset, 80\% of moderator reasons were matched using rule numbers, and the remaining 20\% through similarity-based matching.

% In 80\% of the Lemmy moderation instances in the original Lemmy modlog data, moderators either directly mentioned the rule numbers in their rationales (e.g., ``Rule 1''), or expressed it as a substring of a rule present in the community description (e.g., ``be excellent'', referring to the rule ``Be excellent to each other''). 

\paragraph{Safe comments} To provide a counterbalance to the moderated comments, we also collected a random sample of publicly accessible comments that were not removed by moderators. These are encoded under the ``Safe'' class. To ensure consistency, we stratified this sampling by community, limiting it to the same set of communities that contributed moderated comments, and applied the same filtering criteria used for removed content.

% \usepackage{booktabs}
% \usepackage{graphicx}
\begin{table*}[t]
\centering
\resizebox{\textwidth}{!}{%
\begin{tabular}{@{}rcccccccccccc@{}}
\toprule
 & \textbf{Incivil} & \textbf{Hate} & \textbf{Spam} & \textbf{Content} & \textbf{Doxx} & \textbf{Format} & \textbf{Harass} & \textbf{Meta} & \textbf{Off-tpc} & \textbf{Troll} & \textbf{Other} & \textbf{Safe} \\ \midrule
\textbf{Lemmy} & 8,917 & 4,739 & 2,264 & 6,572 & 618 & 2,217 & 4,202 & 15,304 & 1,703 & 4,412 & 511 & 21,518 \\
\textbf{Reddit} & 9,574 & 815 & 2,574 & 1,655 & 244 & 1,735 & 2,254 & 643 & 1,560 & 843 & - & 31,119 \\ \bottomrule
\end{tabular}%
}
\caption{Rule category distribution in Lemmy and Reddit datasets}
\label{tab:cat_data}
\end{table*}

\subsection{Preparing QA Data}
For each dataset, we retain three key components: the comment text, the full set of community rules (each with an associated rule number), and the specific rule that was violated. We use the violated rule text to derive the model supervision signal. In ModQ-Extract, the violated rule is treated as a span within the concatenated rule set, allowing the model to learn to extract the relevant substring (Figure \ref{fig:introthumb}). In ModQ-Select, the violated rule is instead represented by its corresponding rule number, enabling the model to perform rule selection from a fixed set of candidates (Figure \ref{fig:introthumb}).

%Next, we discuss the proposed, SOTA and the baseline models used in this study. 

\section{Models}

\subsection{ModQ}
We formulate rule prediction as a question-answering (Q\&A) task. Similar to language generation approaches, our method provides the model with a user comment and a contextual input consisting of all applicable community rules. However, instead of asking the model to generate a rule as free-form text, we train it to select the rule that best matches the enforcement decision, either by extracting it as a text span or selecting it from a predefined list.

This reframing allows us to leverage pretrained Q\&A architectures, which are highly effective at answer extraction, while avoiding the verbosity and instability often associated with generative outputs. In addition to being more interpretable, this approach offers practical benefits: by including the full rule set as part of the model’s context at inference time, we enable the system to operate across a variety of communities and rule configurations without retraining. Crucially, our models do not rely on computationally expensive large language models (LLMs), making them suitable for deployment in real-world moderation workflows, including those with limited technical or computational resources.

We implement two model variants aligned with common Q\&A paradigms: an extractive model and a multiple-choice model. Both variants fine-tune the same pretrained transformer backbone—\verb|conversational-bert-base-cased|.\footnote{\url{https://huggingface.co/DeepPavlov/bert-base-cased-conversational}}

\subsubsection{ModQ-Extract}
\paragraph{Model specifications}
ModQ-Extract is an extractive Q\&A model that identifies the span of text within the rule set that corresponds to the rule applied to a given comment. The model is trained to predict the start and end positions of the relevant span using a cross-entropy loss function. To improve stability during optimization, we apply gradient accumulation and exponential moving average.

The model input is a pair \textit{(community, QA)}, where \textit{QA} includes the comment and a concatenated list of all rules in the associated community. The rule list serves as the context for extraction. To help the model learn the structural boundaries between rules, we format each rule as \verb|'{rule_n}. {rule_text}'| and wrap both the comment and the rule context with custom tokens.

At inference time, the model predicts the most likely start and end positions of a span within the concatenated rule context. We then identify which rule contains that span: if the predicted span falls entirely within one rule, that rule is selected; if it overlaps multiple rules, we assign the prediction to the rule that is most fully covered by the span. This design enables fine-grained token-level reasoning while retaining interpretability at the rule level.

\paragraph{Data augmentation}
To encourage generalization and reduce overfitting, we apply three data augmentation strategies during training:
\begin{itemize}
    \item Rule shuffling: Randomizing the order of rules in the context to prevent the model from memorizing positional patterns.
    \item Number permutation: Randomly renumbering rule tokens to prevent the model from overfitting to specific rule indices.
    \item Rule exclusion: Removing one random rule from the context, to prevent the model from memorizing rule sets and enabling it to generalize to their variation. This may include removing the correct rule, forcing the model to learn from counterfactual inputs.
\end{itemize}
%We generate one additional augmented training example per strategy.

\subsubsection{ModQ-Select}
\paragraph{Model specifications}
Unlike extractive Q\&A, multiple-choice Q\&A trains a model to select the correct answer from a set of candidate options. ModQ-Select follows this paradigm by evaluating a user comment against each rule in the community and predicting the likelihood that each rule applies. Similar to the approach in \cite{he2024cpl}, the model scores each comment–rule pair for applicability. However, unlike \cite{he2024cpl}, our model evaluates all candidate rules in a single batch, allowing for efficient and scalable inference over full rule sets.

Formally, the model receives an input sequence of the form \verb|'{content} [SEP] {community}'|, along with the full set of rule texts and the index of the rule enforced by the moderator. During training, the model constructs a separate input for each comment–rule pair and assigns a binary label indicating whether the rule is correct. At inference time, it computes a score for each rule, then selects the rule with the highest predicted score as the final output.

\subsection{State-of-the-Art}
We compare our models against two state-of-the-art baselines: (1) NormVio, a classifier developed for the NormVio dataset that identifies the category of the violated rule (e.g., ``spam''), and (2) CPL-NoViD, a generative, prompt-based approach that extends NormVio by predicting the specific rule being enforced (e.g., ``No excessive reposting'') \cite{park2021detecting,he2024cpl}.

\paragraph{NormVio}

\citeauthor{park2021detecting} model community-sensitive norm violations using transformer-based classifiers that incorporate community-level context. This work serves as a strong baseline for capturing rule-sensitive moderation behavior, particularly on heterogeneous, multi-community platforms like Reddit and Lemmy.
We use NormVio’s best-performing configuration, which takes both the comment text and the community name as input to contextualize rule application. (Notably, more computationally demanding configurations that included additional metadata did not improve performance.) NormVio consists of nine binary classifiers—one per rule category. Unlike our approach, it does not predict the specific rule violated, but rather the broader rule category.
To compare NormVio with other models in our evaluation, we run all category-specific classifiers for each test comment and collect the predicted rule categories. If none of the classifiers are triggered, the comment is labeled as ``safe.''

\paragraph{CPL-NoViD}
\citeauthor{he2024cpl} introduce CPL-NoViD, a prompt-based learning model that incorporates contextual signals such as user history and conversational context to improve accuracy. It represents the state of the art among generative approaches. The model is finetuned using a prompt template of the form:

\textit{``In the [subreddit] subreddit, there is a rule: [rule]. A comment was posted: [comment]. Does the comment violate the subreddit rule? [MASK]''}

For consistency, we use the zero-shot configuration of CPL-NoViD, which excludes prior conversational context. To ensure fair comparison with other models, we evaluate CPL-NoViD on each rule in the community associated with a given test comment, similar to our setup for NormVio.\footnote{The evaluation script provided by the authors tests whether a given comment violates a specific, pre-identified rule but does not compare that rule to others in the same community. That is, if a comment was removed due to "No personal attacks," the model is only tested on that rule—not on whether "No hate speech" might also apply. This limits interpretability at the rule category level and reduces comparability across candidate rules.}

\subsection{Baselines}
%We establish baseline models to evaluate the performance of our proposed approach. 

% \paragraph{Always predict ``safe''} This is the most simplistic strategy, which predicts ``safe'' (the majority class) for all input instances. 
\paragraph{Random predict} This baseline predicts rule applications according to their frequency distribution in the training data. Because the number of rules per comment varies and the rule distribution is highly skewed, this serves as a reference point for comparing model performance against random assignment.

\paragraph{Naive Bayes} We implement a Complement Naive Bayes classifier using TF-IDF vectorization of the comment text. This provides a lightweight and interpretable baseline grounded in traditional text classification techniques.
% \paragraph{Linear SVM} We implement a Linear SVM, transforming text content using Hashing vectorization. 
% \paragraph{BERT} We fine-tune a pre-trained BERT model, with the final layer adapted for multi-class classification, to compare the usefulness of the Q\&A approach over standard classification with the same base model.

We train all baseline models to predict the rule number applied to each comment, treating ``rule 0'' as the class corresponding to safe (i.e., unmoderated) content. One model is trained per community to account for differences in rule sets and moderation patterns across communities. %For the BERT baseline, instead of training one model per community, we incorporate the community identifier as an additional input during the fine-tuning process. This allows the model to learn community-specific representations and potentially improve its ability to predict rule applications within particular community contexts, while benefiting from the larger, more varied combined training data. 
All models' predictions are then aggregated before evaluation.
% \usepackage{booktabs}
% \usepackage{graphicx}
\begin{table*}[]
\centering
\resizebox{0.99\textwidth}{!}{%
\begin{tabular}{@{}rcccccccccccc@{}}
\toprule
 & \multicolumn{1}{l}{\textbf{Incivil}} & \multicolumn{1}{l}{\textbf{Hate}} & \multicolumn{1}{l}{\textbf{Spam}} & \multicolumn{1}{l}{\textbf{Content}} & \multicolumn{1}{l}{\textbf{Doxx}} & \multicolumn{1}{l}{\textbf{Format}} & \multicolumn{1}{l}{\textbf{Harass}} & \multicolumn{1}{l}{\textbf{Meta}} & \multicolumn{1}{l}{\textbf{Off-tpc}} & \multicolumn{1}{l}{\textbf{Troll}} & \multicolumn{1}{l}{\textbf{Other}} & \multicolumn{1}{l}{\textbf{Safe}} \\ \midrule
\textbf{Random} & 0.20 & 0.22 & 0.10 & 0.33 & 0.22 & 0.13 & 0.28 & 0.40 & 0.10 & 0.23 & 0.14 & 0.52 \\
\textbf{Naive bayes} & 0.80 & 0.84 & 0.75 & 0.82 & 0.93 & 0.70 & 0.85 & 0.79 & 0.68 & 0.87 & 0.71 & 0.76 \\ \midrule
\textbf{NormVio} & 0.86 & 0.89 & \textbf{0.85} & 0.87 & 0.93 & 0.75 & \textbf{0.91} & 0.87 & 0.76 & \textbf{0.89} & - & 0.84 \\
\textbf{CPL-NoViD} & \textbf{0.87} & 0.87 & 0.82 & 0.80 & 0.92 & 0.73 & 0.87 & 0.79 & \textbf{0.78} & 0.88 & 0.75 & 0.80 \\ \midrule
\textbf{ModQ-Extract} & 0.85 & 0.88 & 0.82 & 0.86 & 0.96 & 0.73 & 0.88 & 0.86 & 0.77 & 0.88 & 0.71 & 0.85 \\
\textbf{ModQ-Select} & \textbf{0.87} & \textbf{0.90} & 0.84 & \textbf{0.89} & \textbf{0.97} & \textbf{0.76} & \textbf{0.91} & \textbf{0.88} & \textbf{0.78} & \textbf{0.89} & \textbf{0.77} & \textbf{0.87} \\ \bottomrule
\end{tabular}%
}
\caption{\textbf{Lemmy ModQ macro F1 results: }Table providing comparative macro F1 results on the Lemmy dataset aggregated across categories between different models. Our proposed models---ModQ variants---consistently outperform other models across most categories.}
\label{tab:lemmy_test_stratified}
\end{table*}

\section{Experimental Set-Up}
% \usepackage{booktabs}
% \usepackage{graphicx}
\begin{table*}[]
\centering
\resizebox{0.88\textwidth}{!}{%
\begin{tabular}{@{}rccccccccccc@{}}
\toprule
 & \multicolumn{1}{l}{\textbf{Incivil}} & \multicolumn{1}{l}{\textbf{Hate}} & \multicolumn{1}{l}{\textbf{Spam}} & \multicolumn{1}{l}{\textbf{Content}} & \multicolumn{1}{l}{\textbf{Doxx}} & \multicolumn{1}{l}{\textbf{Format}} & \multicolumn{1}{l}{\textbf{Harass}} & \multicolumn{1}{l}{\textbf{Meta}} & \multicolumn{1}{l}{\textbf{Off-tpc}} & \multicolumn{1}{l}{\textbf{Troll}} & \multicolumn{1}{l}{\textbf{Safe}} \\ \midrule
\textbf{Random} & 0.49 & 0.49 & 0.49 & 0.49 & 0.49 & 0.49 & 0.49 & 0.49 & 0.49 & 0.49 & 0.28 \\
\textbf{Naive bayes} & 0.44 & 0.49 & 0.48 & 0.49 & 0.49 & 0.49 & 0.48 & 0.49 & 0.49 & 0.49 & 0.28 \\ \midrule
\textbf{NormVio} & 0.85 & 0.76 & 0.81 & 0.74 & 0.66 & 0.80 & 0.85 & 0.81 & 0.74 & 0.72 & 0.85 \\
\textbf{CPL-NoViD} & 0.72 & 0.79 & 0.51 & 0.82 & \textbf{0.91} & 0.54 & 0.85 & 0.54 & 0.78 & 0.80 & 0.53 \\ \midrule
\textbf{ModQ-Extract} & 0.85 & \textbf{0.82} & 0.85 & 0.77 & 0.81 & 0.82 & 0.87 & \textbf{0.83} & \textbf{0.81} & \textbf{0.81} & 0.83 \\
\textbf{ModQ-Select} & \textbf{0.86} & 0.81 & \textbf{0.86} & \textbf{0.79} & 0.76 & \textbf{0.84} & \textbf{0.89} & \textbf{0.83} & 0.80 & 0.80 & \textbf{0.86} \\ \bottomrule
\end{tabular}%
}
\caption{\textbf{Reddit ModQ macro F1 results: }Table providing comparative macro F1 results on the Reddit dataset aggregated across categories between different models. Our proposed models---ModQ variants---consistently outperform other models across most categories.}
\label{tab:reddit_test}
\end{table*}

\paragraph{Data splits}
To rigorously evaluate model performance, we employed a multi-stage data splitting strategy. We created train, development, and test sets through random sampling respectively 80\%, 10\%, and 10\% of the data, stratifying the proportions of moderated comments in each split.   

Prior to creating these splits, we held out data from two sources to simulate real-world generalization scenarios. First, we excluded all data from a random sample of $N = 20$ communities to form a \textit{leave-N-communities-out} test set. Second, we excluded $N = 20$ randomly selected rules—defined as unique (rule, community) pairs—from the remaining dataset to construct a \textit{leave-N-rules-out} test set.

These two held-out test sets present challenging out-of-distribution conditions, simulating practical settings where new communities are created or where new rules get added.

\paragraph{Training}
We train both ModQ-Extract and ModQ-Select models for 5 epochs, selecting the checkpoint with the highest macro F1 score on the development set for final evaluation. Both models are trained using a learning rate of $1 \times 10^{-5}$ and a weight decay of $0.001$. Preliminary experiments with alternative learning rates and weight decay values yielded comparable results. For state-of-the-art baselines, we retain the original training configurations provided by the respective authors. All models were trained on an NVIDIA L40 GPU, with average per epoch training time between 18 to 35 minutes depending on the model. ModQ-Select model was also trained parallelly on RTX 3060 with 1 hour per epoch time.

\paragraph{Evaluation} We report performance over rule categories, which summarize how well models perform on common classes of rule transgressions. We associate each comment with the ground-truth and predicted categories of the rules it was moderated for, adding the ``safe'' category to comments that either were safe in the ground-truth, or for which models did not predict any other category. To enable comparison with the SOTA, we treat each category as a separate binary classification task and macro-average performance between the two classes (category, not-category). %To better represent the multi-label nature of the task and the ability of the model to identify each category, we also report per-category metrics. For example, a classifier may identify comments moderated for spam at $F1=0.5$ (per-category metric) and those not moderated for spam at $F1=1$, leading to an overall $F1=0.75$ (macro-average metrics). Since the majority class is the one for which each category does not apply, per-category metrics are more conservative estimates of model performance. 
We also provide binary classification metrics distinguishing whether a comment violates \emph{any} moderation rule or is ``safe'', which allows clearer inspection of the model’s safety detection.

\section{Results}

Next, we describe the results of model evaluation. We start by discussing overall performance ---
Tables \ref{tab:reddit_test} and \ref{tab:lemmy_test_stratified} summarize key metrics of model performance on the Reddit and Lemmy datasets, respectively. We break down performance according to categories of transgressions and the binary task of determining whether a comment should be moderated. 
We then unpack how performance translates to challenges in real-world settings. We study to what extent models generalize to new rules and communities. Finally, we discuss which rule categories are often conflated by the model, assessing to what extent erroneous predictions could mislead moderators.
Note that the performance breakdown by categories is only performed to offer direct comparison with other models. 

\subsection{Performance on Different Rule Categories}
Both ModQ variants demonstrate strong and competitive performance relative to state-of-the-art (SOTA) models on the Reddit and Lemmy datasets. Across nearly all rule categories, ModQ-Select consistently outperforms all other models, including its extractive counterpart. ModQ-Extract also achieves performance on par with or just slightly below the best-performing baselines, despite tackling the more challenging task of identifying the specific rule text rather than a broad category.

On Lemmy, ModQ-Select reaches the highest F1 scores in nearly every category, including difficult-to-detect ones like content ($0.89$), harassment ($0.91$), and doxxing ($0.97$). ModQ-Extract closely follows, maintaining F1 scores comparable to NormVio and CPL-NoViD, despite using a single model rather than multiple binary classifiers. This underscores the strength of the extractive QA formulation in retaining interpretability while matching state-of-the-art performance.

\begin{figure*}[t]
    \centering
    \includegraphics[width=0.99\textwidth]{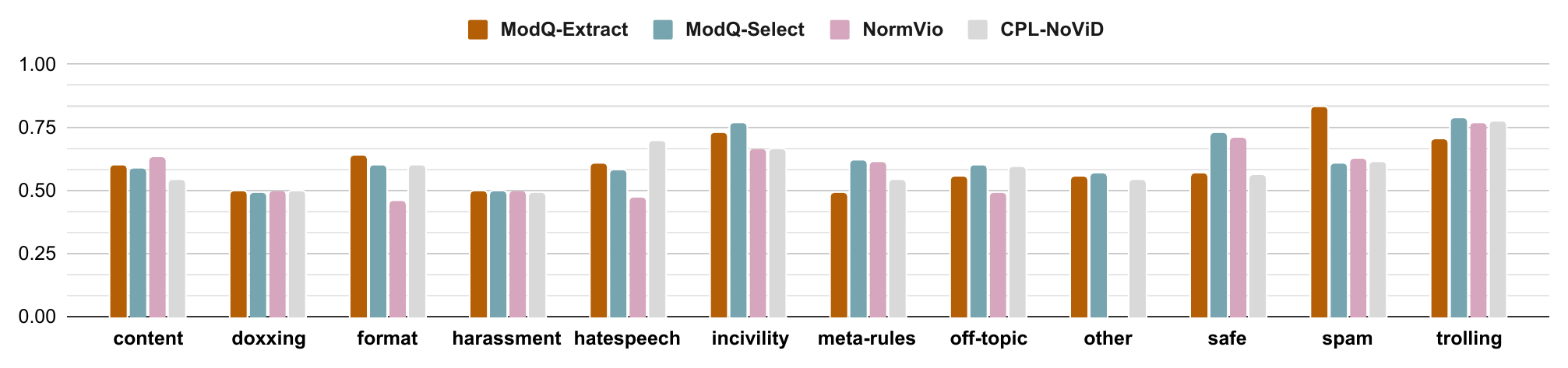}
    \caption{\textbf{Lemmy data \textit{leave-N-communities-out}: }Figure illustrating macro F1 results for \textit{leave-N-communities-out} test set.}
    \label{fig:n_communities_out}
\end{figure*}

\begin{figure*}[t]
    \centering
    \includegraphics[width=0.65\textwidth]{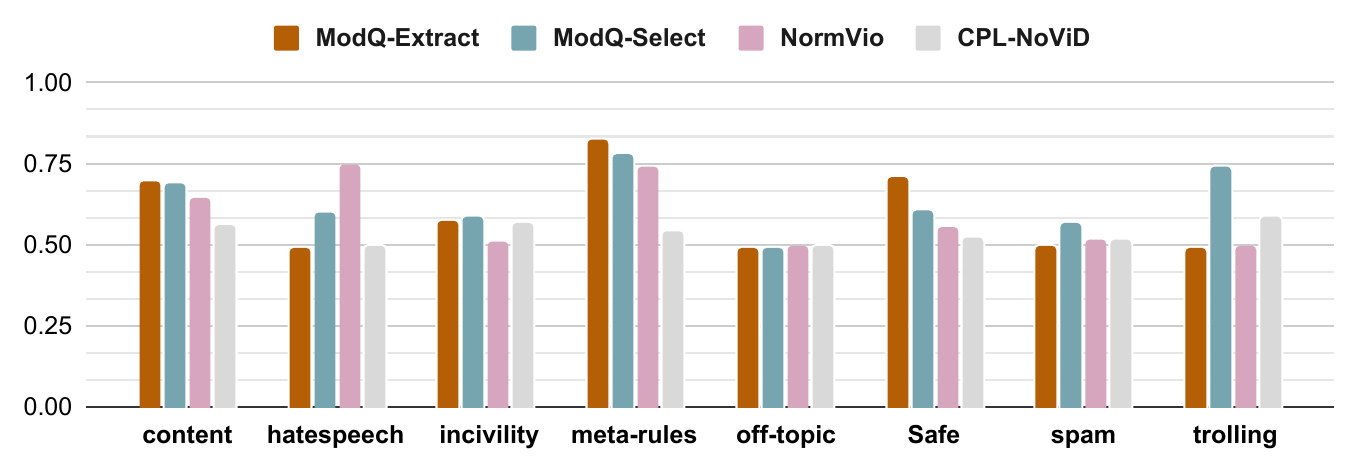}
    \caption{\textbf{Lemmy data \textit{leave-N-rules-out}: }Figure illustrating macro F1 results for \textit{leave-N-rules-out} test set.}
    \label{fig:n_rules_out}
\end{figure*}

On Reddit, ModQ-Select again outperforms other models across most categories. For instance, it surpasses CPL-NoViD in incivility, spam, and harassment, and matches or exceeds NormVio in safe and trolling categories. Notably, both ModQ variants outperform CPL-NoViD on ``safe'' classification—a critical measure for avoiding false positives in moderation. Thus, both ModQ variants offer strong performance on the  NormVio dataset, while additionally providing granular, per-rule predictions akin to CPL-NoViD. Notably, ModQ-Select achieves this strong performance with lower computational demands than both ModQ-Extract and CPL-NoViD.

Overall, these results demonstrate that our proposed ModQ models offer a favorable balance of accuracy, interpretability, and efficiency—achieving strong alignment with moderator decisions while delivering more fine-grained rule-level outputs than prior work.

% \usepackage{booktabs}
% \usepackage{graphicx}
\begin{table}[]
\centering
\resizebox{\columnwidth}{!}{%
\begin{tabular}{@{}lcc@{}}
\toprule
 & \textbf{Safe  (macro F1)} & \multicolumn{1}{l}{\textbf{Not Safe (macro F1)}} \\ \midrule
\textbf{NormVio} & 0.85 & 0.83 \\ \midrule
%\textbf{CPL-NoViD} & - & - \\ \midrule
\textbf{ModQ-Extract} & 0.86 & 0.84 \\
\textbf{ModQ-Select} & \textbf{0.87} & \textbf{0.86} \\ \bottomrule
\end{tabular}%
}
\caption{Binary classification macro F1 for the Lemmy dataset. ModQ-Select outperforms all other models on the binary task. }
\label{tab:binary}
\end{table}

\subsection{Binary rule classification task}
% \subsection{Alignment with Moderation Decisions}
% Both ModQ variants outperform the SOTA on the binary task of predicting whether a comment should be approved or removed according to moderation guidelines, with TODO and TODO F1 respectively on Reddit, and TODO and TODO on Lemmy. TODO CHECK THE FOLLOWING IS TRUE Both models in both datasets prioritize precision over recall, corresponding to fewer cases of overcensorship than of undermoderation. This improved alignment with actual moderation decisions demonstrates the practical applicability of this approach. 

Both ModQ variants outperform the SOTA on the binary task of predicting whether a comment should be approved or removed according to moderation guidelines. For NormVio task we considered a comment to be 'Safe' if it is not flagged any of the classifiers. CPL-NoViD setup is inherently incompatible with binary safe-not safe prediction task \cite{he2024cpl}. Hence, we do not include CPL-NoViD results to avoid unfair comparison.

\subsection{Generalization to Unseen Rules and Communities}

Generalization to unseen communities and rules remains a significant challenge across all models. As expected, performance decreases when models are evaluated on out-of-distribution settings—either due to new community contexts (leave-N-communities-out) in Figure \ref{fig:n_communities_out} or novel rule sets (leave-N-rules-out) in Figure \ref{fig:n_rules_out}. These conditions reflect real-world scenarios where new communities emerge or moderation policies evolve, demanding robustness beyond static, in-domain training.

Despite this performance drop, both ModQ-Extract and ModQ-Select consistently outperform strong baselines, including NormVio and CPL-NoViD, across nearly all categories in these challenging settings. In the leave-N-communities-out test (Figure~\ref{fig:n_communities_out}), ModQ-Select leads across categories, particularly in incivility, off-topic, meta-rules, safe, and trolling, suggesting its strength in transferring moderation behavior across different community norms. ModQ-Extract performs competitively as well, with only marginally lower performance than ModQ-Select, and often exceeding other models in format, and spam. 

In the leave-N-rules-out setting (Figure~\ref{fig:n_rules_out}), where models must generalize to previously unseen moderation policies, the gap between our models and the baselines becomes even more evident. Both ModQ variants maintain stronger macro F1 scores across categories like content, incivility, meta-rules, safe, trolling and spam. This suggests that ModQ’s QA-based formulation effectively leverages the textual structure of rules, enabling it to handle semantic variation in rule wording better than other models.

Taken together, these results demonstrate that ModQ models not only match or exceed state-of-the-art performance in-distribution but also offer superior robustness in out of distribution tests. This capability is particularly promising for bootstrapping moderation systems in new or low-resource communities, where labeled training data is scarce or non-existent. By conditioning on the rule set at inference time, rather than relying on fixed taxonomies or extensive fine-tuning, ModQ supports adaptive, scalable moderation that aligns with evolving community standards.

\begin{figure*}[t!]
    \centering
    \includegraphics[width=0.95\linewidth]{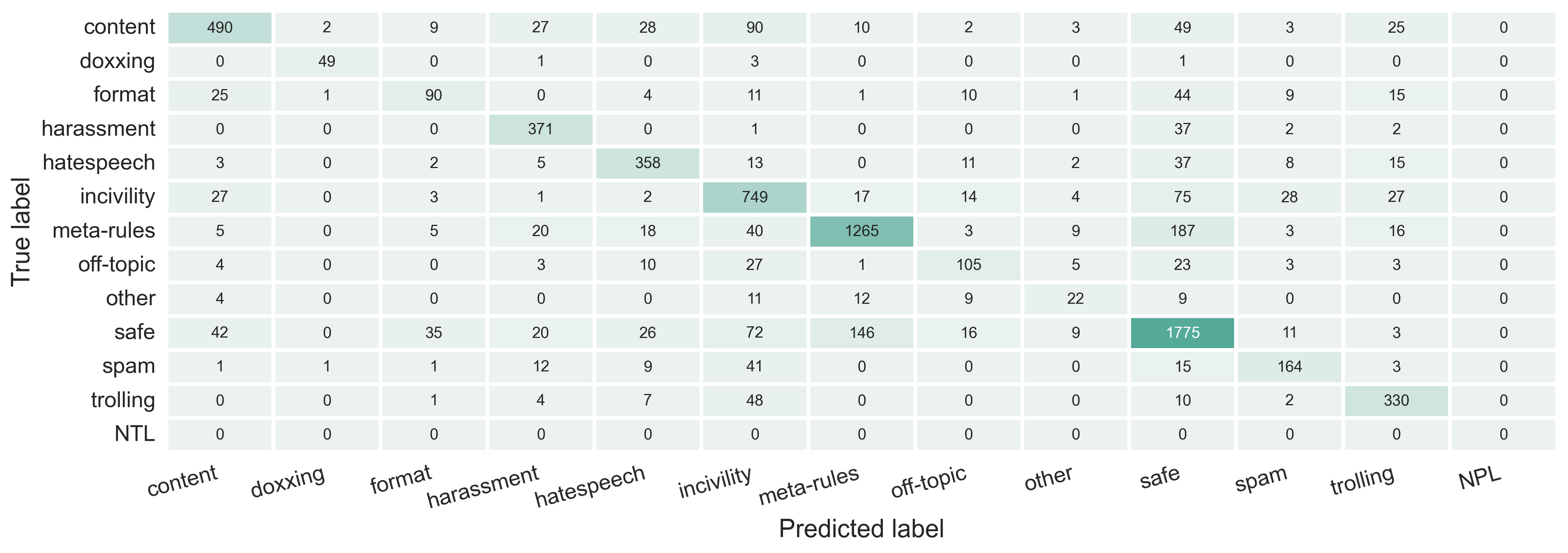}
    \caption{Confusion matrix produced by ModQ-Select between true (y axis) and predicted (x axis) labels}
    \label{fig:confusion}
\end{figure*}
% \subsection{Applicability of ModQ}
% Focusing on the best-performing model, ModQ-Select, we unpack how it could be applied in online communities. To this end, first, we examine when it successfully predicts individual rules, and on average, how performance fluctuates between communities. Then, we discuss which rule categories are often conflated by the model, assessing to what extent erroneous predictions could mislead moderators.

% \subsection{Performance across Rules and Communities}
% - top rules are ...; why?
%     - confounder: support (fewer instances to learn from) @sp watch out, df has data on test but there may be a mismatch with train distributions. ideally, we could gett support from train and match with performance on test
%     - confounder: index (most important rules come first, even if we try to disentangle rule positions)
%     - scatter plot or reg plot: support (log scaled) vs f1
% - top communities are ...; why
%     - confounder: length of the rule set (many options to choose from, harder to predict)
%     - confounder: community size?
%     - scatter plot or reg plot: number of options vs f1; one curve with the random probability of the rule

\subsection{Confusion between Rule Categories}

% - We report Multi-Label Confusion Matrices as detailed in \cite{mlcm} to show relationships between ground-truth and predicted categories. This allows us to unpack how not only which infraction categories are mislabeled by the model, but also how they would have been reported.
% - model makes reasonable errors: 
%     - overpredicts safe (better to be lenient than to censor) @sp in case we use this argument, remove it from the general performance section
%     - confuses semantically related infraction categories, 
%         - meta rules include behavior control including incivility, harassment, trolling
%         - content, format, and spam can be similar
%     - some categories are hard/don't have enough data. works surprisingly well on e.g., doxxing despite this.
%     - pull examples of rules!
To better understand model behavior and failure modes, we analyze which rule categories are most frequently confused with one another. For brevity, we focus on the best-performing model—ModQ-Select—and visualize its predictions in the confusion matrix shown in Figure~\ref{fig:confusion}.\footnote{We report a Multi-Label Confusion Matrix, as each rule is associated with one or more categories \cite{mlcm}.}

A common pattern is the overprediction of the “safe” label, particularly when the true category belongs to meta-rules, incivility, content, or format. These categories often require contextual or interpretive nuance, and moderation boundaries may be blurry. Importantly, this overprediction skews the model toward caution—minimizing false positives at the cost of some false negatives—making it less likely to flag borderline cases, which may be desirable in human-in-the-loop moderation settings.

Beyond “safe” misclassifications, the model's errors are generally semantically reasonable. For instance
%, incivility is often confused with harassment, reflecting overlap in behavioral expectations (e.g., “Keep it civil” vs. “Don’t be insulting”). 
format is often predicted as content as the two categories are often entangled  by governing post structure and informational quality (e.g., “Mark spoilers” [content], “Tag all posts” [format]. Furthermore, meta-rules, which often include behavioral expectations not tied to a single post, are mistaken for incivility, likely due to their conceptual breadth and lack of precise linguistic markers.

Some categories remain particularly difficult due to abstractness or low support. For example, meta-rules, off-topic, and other violations often lack consistent textual cues, making them harder to learn. Conversely, doxxing—despite being rare—is detected with high reliability, likely due to its more distinctive linguistic signatures and narrow scope.

Overall, this analysis suggests that ModQ-Select’s errors tend to arise in borderline or ambiguous cases, particularly where rule boundaries are inherently soft or where training data is sparse. This reinforces the importance of using such models as decision support tools rather than fully autonomous moderators, particularly in nuanced domains.

%Moreover, the model's error patterns are generally associated with semantically similar categories. For example, ``incivility'' and ``harassment'', ``Keep it civil'' (incivility) and ``Don’t be abrasive/insulting'' (harassment) are rules that may be both applicable to similar comments and be interpreted similarly. By the same token, violations to ``format'', ``content'', and ``spam'' rules all involve text structure or quality, making them difficult to distinguish, e.g., ```All posts must be tagged'' (format), `Mark spoilers'' (content) and ``No spam or promotional material'' (spam).

%Some categories remain inherently challenging, arguably due to abstract definitions. For example, ``meta-rules'', ``off-topic'', and ``other'' violations---lacking clear, general textual signatures---are frequently confused. While ``doxxing'' is detected with surprising reliability despite being rare, the performance for other categories generally improves with proportionally larger representation in the training data.

\section{Discussion and Implications}

Before concluding, we discuss implications of our work for content moderation practices.

\subsection{Moderation Support}
% - moderation support
%     flagging system for content
%     rule hint for decision
%     rationale autocomplete
One of the key contributions of this work lies in its potential to enhance the design of moderation support tools. By producing granular, per-rule predictions, our models improve both the interpretability and transparency of automated systems—two features that current machine learning-based moderation tools often lack. Rather than reducing decisions to opaque binary outcomes, the ModQ framework provides actionable outputs tied directly to community-authored rules. This capability unlocks several practical applications. First, it can power flagging systems that not only identify potentially violating content but also specify the rule it may contravene. This helps moderators make faster and more informed decisions, particularly in high-volume environments where consistency is challenging. Further, ModQ’s outputs can be used to automatically draft rationale templates for enforcement actions. Rather than requiring moderators to manually justify each action---a known source of cognitive and emotional labor---automated rationales grounded in the rule set can streamline communication with users, while also promoting consistency and accountability. In community contexts where moderation actions are public, this can foster greater trust and reduce perceptions of arbitrariness. Finally, by surfacing rule-based predictions proactively, ModQ can support user-facing nudging tools. For example, when a user begins composing a comment, the model could flag likely rule violations and suggest revisions before submission—much like spell-check or grammar-correction tools. Such preemptive interventions may reduce downstream conflict, lighten the load on moderators, and improve the overall health of online communities.

%Providing granular, per-rule predictions may improve the transparency and interpretability of ML-powered moderation support tools, which have been identified as a crucial shortcoming of current solutions. The proposed framework can support the development of such tools, including flagging systems that highlight potentially violating content and suggest the most relevant rules. Similarly, the model could be leveraged for drafting rationales for moderator actions, as well as providing proactive hints to users before posting online. Overall, Q\&A-based tools may provide users, moderators, and community managers with crucial context about the application of community guidelines.

\subsection{Low-Resource Moderation}
% - low-resource moderation 
%     data-poor new communities leveraging pre-trained model
%     compute-poor communities pooling together a shared hosted model

Our approach holds particular promise for addressing the challenges of content moderation in low-resource settings. Notably, ModQ's low computational demands compared to the SOTA make it an accessible solution for computationally-versed community moderators. New online communities, often lacking the extensive moderation logs required to train conventional supervised models, can leverage Q\&A models pre-trained on other communities to bootstrap their moderation efforts. Furthermore, ModQ's adaptability to different rule sets at inference time implies that communities with limited computational resources can benefit from sharing models or accessing third-party-hosted ones, which they can access on demand. 

\subsection{Governance Insights}
% - governance insights
%     test futures/scenarios with different rule sets
%     alignment?

Beyond applications in content moderation, modeling the relationship between rules and their enforcement at scale enables quantifying and exploring crucial aspects of governance in online communities. The Q\&A formulation allows us to treat community rules as a structured knowledge base, which can then be queried to study their role in community health. For example, the model can be used to test hypothetical moderation scenarios under different rule sets, enabling communities to proactively evaluate the impact of proposed rule changes. Moreover, model outputs can be analyzed to assess the alignment between formal rules and actual moderation decisions. Discrepancies between predicted rule matches and historical moderation behavior can reveal enforcement gaps, inconsistent application of norms, or implicit biases in moderator judgments. These insights may inform community audits, deliberative reform processes, or broader questions of legitimacy and transparency in digital governance. ModQ provides not just a tool for rule enforcement, but a lens through which to study the institutional logic of rule-governed systems online. 
%Furthermore, the model's predictions can be analyzed to assess the alignment between stated rules and actual moderation practices, providing valuable insights into the effectiveness and transparency of community governance.

\section{Limitations and Ethical Considerations}
% - imperfect data; esp. reddit. not all rules on reddit; truncated rule texts; although followed best practices and evaluated along the way, the data processing includes several inference steps, e.g.,  for rule extraction, rule matching, rule categorization, which carry inherent risks for inaccuracy
% - acknowledge potential bias; especially, only english; 
% - warn that is good but not perfect; sometimes, even moderators disagree: ground truth may not exist; should be used as flagging system and not autonomous
% - moderation data is sensitive data; especially, modlogs on lemmy carry risk of revictimization, brigading against transgressors; retaliation against moderators;

Several limitations in our data and methodology warrant discussion. First, the replication of NormVio and CPL-NoViD, and the testing of ModQ on the Reddit dataset are conditioned by missing data, due to the restriction of Reddit's API access. Especially, we could not retrieve all subreddit rules to make a comparable setup as for Lemmy; we could also not rehydrate the text of the rules that are truncated in the original dataset. More broadly, although we followed best practices and evaluated each step, the data processing pipeline includes several inference steps (e.g., for rule extraction, rule matching, and rule categorization), each of which carries inherent risks of inaccuracy.

We also acknowledge the potential for bias in our dataset, starting from its limitation to English-language communities. While our approach shows promising performance, it is important to emphasize that content moderation is a broader and more complex task, involving different roles and expertise, and every transgression and normative act requires contextualization. Notably, a definitive ``ground truth'' about what should be moderated and whether a rule applies may always exist---even expert human moderator teams disagree on whether a comment violates a rule. Therefore, the models we propose are intended as supports to scale up the effort of human moderators, for example, as a flagging system, rather than as an autonomous decision-making tool.

Finally, we recognize that moderation data is sensitive. In particular, the moderator logs we collected from Lemmy carry the risk of re-victimization of the targets of moderated comments, brigading against those whose content was moderated, as well as potential retaliation against moderators. Although this data is publicly accessible through Lemmy API, we have decided not to share to limit these potential harms.

\section{Conclusions}
This paper introduced Q\&A as a novel formulation of automated content moderation in online communities, and introduced ModQ: two model variants that identify whether a comment infringes community rules. The proposed models correspond to extractive and multiple-choice Q\&A tasks. The models, although computationally lightweight, outperform the state of the art in emulating approve/remove moderation decisions as well as in detecting content across a range of common infraction categories. This framework applies out-of-the-box to rule sets that may vary from community to community or over time.

\bigskip
%\noindent Thank you for reading these instructions carefully. We look forward to receiving your electronic files!

\bibliography{aaai25}
% \appendix
% \section{Reference Examples}
% \label{sec:reference_examples}

\subsection{Paper Checklist to be included in your paper}

\begin{enumerate}

\item For most authors...
\begin{enumerate}
    \item  Would answering this research question advance science without violating social contracts, such as violating privacy norms, perpetuating unfair profiling, exacerbating the socio-economic divide, or implying disrespect to societies or cultures?
    \answerYes{Yes}
  \item Do your main claims in the abstract and introduction accurately reflect the paper's contributions and scope?
    \answerYes{Yes}
   \item Do you clarify how the proposed methodological approach is appropriate for the claims made? 
    \answerYes{Yes}
   \item Do you clarify what are possible artifacts in the data used, given population-specific distributions?
    \answerYes{Yes}
  \item Did you describe the limitations of your work?
    \answerYes{Yes}
  \item Did you discuss any potential negative societal impacts of your work?
    \answerYes{Yes}
      \item Did you discuss any potential misuse of your work?
    \answerYes{Yes}
    \item Did you describe steps taken to prevent or mitigate potential negative outcomes of the research, such as data and model documentation, data anonymization, responsible release, access control, and the reproducibility of findings?
    \answerYes{Yes}
  \item Have you read the ethics review guidelines and ensured that your paper conforms to them?
    \answerYes{Yes}
\end{enumerate}

\item Additionally, if your study involves hypotheses testing...
\begin{enumerate}
  \item Did you clearly state the assumptions underlying all theoretical results?
    \answerNA{NA}
  \item Have you provided justifications for all theoretical results?
   \answerNA{NA}
  \item Did you discuss competing hypotheses or theories that might challenge or complement your theoretical results?
    \answerNA{NA}
  \item Have you considered alternative mechanisms or explanations that might account for the same outcomes observed in your study?
    \answerNA{NA}
  \item Did you address potential biases or limitations in your theoretical framework?
    \answerNA{NA}
  \item Have you related your theoretical results to the existing literature in social science?
    \answerNA{NA}
  \item Did you discuss the implications of your theoretical results for policy, practice, or further research in the social science domain?
   \answerNA{NA}
\end{enumerate}

\item Additionally, if you are including theoretical proofs...
\begin{enumerate}
  \item Did you state the full set of assumptions of all theoretical results?
    \answerNA{NA}
	\item Did you include complete proofs of all theoretical results?
    \answerNA{NA}
\end{enumerate}

\item Additionally, if you ran machine learning experiments...
\begin{enumerate}
  \item Did you include the code, data, and instructions needed to reproduce the main experimental results (either in the supplemental material or as a URL)?
    \answerNo{No, to preserve anonymity at the double blind review stage. We will release code and instructions upon acceptance, but not the data for ethical considerations.}
  \item Did you specify all the training details (e.g., data splits, hyperparameters, how they were chosen)?
   \answerYes{Yes}
     \item Did you report error bars (e.g., with respect to the random seed after running experiments multiple times)?
    \answerNo{No, because we observed very small variability over multiple runs of different splits}
	\item Did you include the total amount of compute and the type of resources used (e.g., type of GPUs, internal cluster, or cloud provider)?
    \answerYes{Yes}
     \item Do you justify how the proposed evaluation is sufficient and appropriate to the claims made? 
   \answerYes{Yes}
     \item Do you discuss what is ``the cost`` of misclassification and fault (in)tolerance?
    \answerYes{Yes}
  
\end{enumerate}

\item Additionally, if you are using existing assets (e.g., code, data, models) or curating/releasing new assets, \textbf{without compromising anonymity}...
\begin{enumerate}
  \item If your work uses existing assets, did you cite the creators?
    \answerYes{Yes}
  \item Did you mention the license of the assets?
    \answerNA{NA}
  \item Did you include any new assets in the supplemental material or as a URL?
    \answerNA{NA}
  \item Did you discuss whether and how consent was obtained from people whose data you're using/curating?
    \answerYes{Lemmy data collected is publicly available through Lemmy API. }
  \item Did you discuss whether the data you are using/curating contains personally identifiable information or offensive content?
    \answerYes{Yes, and we are not releasing this dataset for ethical considerations}
\item If you are curating or releasing new datasets, did you discuss how you intend to make your datasets FAIR (see \citet{fair})?
\answerYes{We are not releasing this dataset for ethical considerations}
\item If you are curating or releasing new datasets, did you create a Datasheet for the Dataset (see \citet{gebru2021datasheets})? 
\answerNA{NA}
\end{enumerate}

\item Additionally, if you used crowdsourcing or conducted research with human subjects, \textbf{without compromising anonymity}...
\begin{enumerate}
  \item Did you include the full text of instructions given to participants and screenshots?
    \answerNA{NA}
  \item Did you describe any potential participant risks, with mentions of Institutional Review Board (IRB) approvals?
    \answerNA{NA}
  \item Did you include the estimated hourly wage paid to participants and the total amount spent on participant compensation?
    \answerNA{NA}
   \item Did you discuss how data is stored, shared, and deidentified?
   \answerNA{NA}
\end{enumerate}

\end{enumerate}

\end{document}